\documentclass[12pt]{article}
\usepackage{amssymb}
\date{}
\newtheorem{theorem}{Theorem}

\newcommand{\bbeth}{{\bar\eth}}

\newcommand{\lam}{{\lambda}}
\newcommand{\blam}{{\bar\lambda}}
\newcommand{\De}{{D'}}

\newcommand{\bA}{{\bar A}}
\newcommand{\al}{\alpha}
\newcommand{\bal}{{\bar\alpha}}
\newcommand{\be}{\beta}
\newcommand{\bbe}{{\bar\beta}}
\newcommand{\de}{\delta}
\newcommand{\bde}{{\bar\delta}}

\newcommand{\bm}{{\bar m}}

\newcommand{\sig}{{\sigma}}
\newcommand{\bga}{{\bar\gamma}}

\newcommand{\btau}{{\bar\tau}}
\newcommand{\bmu}{{\bar\mu}}
\newcommand{\bnu}{{\bar\nu}}

\newcommand{\brho}{{\bar\rho}}
\newcommand{\bPsi}{{\bar\Psi}}
\newcommand{\bph}{{\bar\phi}}

\newcommand{\bze}{{\bar\zeta}}

\newcommand{\ppr}{{\frac{\partial}{\partial r}}}

\newcommand{\ppu}{{\frac{\partial}{\partial u}}}
\newcommand{\ppz}{{\frac{\partial}{\partial\zeta}}}
\newcommand{\ppbz}{{\frac{\partial}{\partial{\bar\zeta}}}}
\newcommand{\bee}{\begin{eqnarray}}
\newcommand{\ede}{\end{eqnarray}}
\newcommand{\eqs}[1]{Eqs. (\ref{#1})}
\newcommand{\eq}[1]{Eq. (\ref{#1})}
\newcommand{\meq}[1]{(\ref{#1})}
\begin{document}
\baselineskip=14 pt
\begin{center}
{\Large\bf On Newman-Penrose constants of stationary electrovacuum
spacetimes }
\end{center}
\begin{center}
Xiangdong Zhang${}^{a,}$\footnote{e-mail :
zhangxiangdong@mail.bnu.edu.cn } Xiaoning
Wu${}^{b,}$\footnote{e-mail : wuxn@amss.ac.cn} and Sijie Gao
${}^{a,}$\footnote{e-mail : sijie@bnu.edu.cn}
\end{center}
\begin{center}
a. Department of Physics, \\
Beijing Normal University,\\
Beijing, China, 100080.\\
b. Institute of Applied Mathematics. \\Academy of Mathematics and
System Science.\\Chinese Academy of Sciences,\\
P.O.Box 2734, Beijing, China, 100080.\\
\end{center}
\begin{abstract}
A theorem related to the Newman-Penrose constants is proven. The
theorem states that  all the Newman-Penrose constants of
asymptotically flat, stationary, asymptotically algebraically
special electrovacuum spacetimes are zero. Straightforward
application of this theorem shows that all the Newman-Penrose
constants of the Kerr-Newman spacetime must vanish.
\end{abstract}

PACS number : 04.20.-q, 04.20.Ha

Keywords : Newman-Penrose constants, stationary electrovacuum
condition, Kerr-Newman solution.

\section{Introduction}

Newmen-Penrose(N-P) constants are very interesting and useful
quantities in the study of asymptotic flat space-times. They were
first found by E.T.Newman and R.Penrose in 1968\cite{NP68} and then
discussed by many other
authors\cite{Price,Kr99,Kr00,FK00,Bo01,Kr02}. Although the N-P
constants have been found for forty years, their physical
interpretation remains an open question. One reason is that the
computation of  these constants for a general asymptotically flat
spacetime is not easy. In stationary vacuum cases, these constants
can be viewed as combination of multi-pole moments of
space-times\cite{Ger70}-\cite{Fr06}. Calculations of the NP
constants for vacuum solutions have been made by many authors
\cite{Ki,LK00,DK02,Wu07,Bai06,Kr07}. People used to conjecture that
the algebraically special condition (ASC) leads to the vanishing of
NP constants. However, Kinnersley and Walker\cite{Ki} provided a
counterexample. Recently, some authors\cite{Wu07} proposed the
asymptotically algebraically special condition (AASC),  and proved
that the N-P constants vanish for vacuum, stationary, asymptotic
algebraically special space-times. In fact, the two conditions are
closely related. It is well known that the ASC implies that the Weyl
curvature possesses a multiple principle null direction. This
condition can be expressed in terms of two geometric invariants $I$
and $J$, defined by $I=\Psi_0\Psi_4-4\Psi_1\Psi_3+3(\Psi_2)^2$ and
$J=\Psi_4\Psi_2\Psi_0+2\Psi_3\Psi_2\Psi_1-(\Psi_2)^3-(\Psi_3)^2\Psi_0-(\Psi_1)^2\Psi_4$
\cite{PR86,Kr80}. A spacetime is said to be algebraically special if
$I^3-27J^2=0$. It has been shown that a general asymptotically flat,
stationary spacetime satisfies $I^3-27J^2\sim O(r^{-21})$ near
future null infinity\cite{Wu07}. Thus, $I^3-27J^2$ will peel off
very quickly for a general stationary vacuum asymptotically flat
spacetime although the spacetime may not be algebraically special. A
spacetime is said to be ``asymptotically algebraically special"  if
$I^3-27J^2\sim O(r^{-22})$ \cite{Wu07}, i.e. one order faster than
general cases.  From the geometric point of view, this indicates
that one pair of principle null directions coincides near null
infinity. By imposing this condition, authors of
\cite{Wu07} showed that the NP constants vanish for vacuum,
stationary spacetimes. Based on the result of \cite{Wu07}, N-P
constants can bee seen as combination of Janis-Newman multi-poles of
gravitational field\cite{JN65}. An intriguing question is  whether
the Janis-Newman multi-poles of matter field will contribute to the
N-P constants. In this paper, we extend the discussion to the
electrovacuum case. By imposing the AACS, we show that the NP
constants still vanish in the presence of a stationary Maxwell
field. If the Maxwell field is not stationary, the multi-pole
moments of the Maxwell field will contribute to the N-P constants.

This paper is organized as follows : In section II, we apply the
method of Taylor expansion to a stationary electrovacuum space-time.
With the help of the Killing equation, we reduce the dynamical
freedom of gravitational field into a set of arbitrary constants.
Detailed expressions are given up to order $O(r^{-6})$. We then
prove that all the N-P constants of a stationary asymptotically
algebraically special electrovacuum space-time are zero. Finally, we
make some concluding remarks in section III.

\section{The Newman-Penrose constants of stationary asymptotically
algebraically special electrovacuum space-times}

In an asymptotically flat spacetime, the Newman-Penrose constants
are defined by  \cite{PR86}
\begin{eqnarray}
G_m=\int_{S_{\infty}}{}_2Y_{2,m}\Psi^1_0 dS,\nonumber
\end{eqnarray}
where ${}_2Y_{2,m}$ is a spin-weight harmonic function and $\Psi^1_0
$ is a component of the Weyl tensor. Since the integral is performed
on a two-sphere at infinity, we only need the asymptotic form of the
Weyl tensor in the calculation. According to the peeling off theorem
given by Sachs \cite{PR86}, we may express the Weyl tensor and
Maxwell field as :
\begin{eqnarray}
&&\Psi_n\sim O(r^{n-5})\, \quad n=0,1,2,3,4,\nonumber\\
&&\phi_m\sim O(r^{m-3})\quad m=0,1,2.
\end{eqnarray}
The vacuum case has been studied
previously\cite{Kr02,LK00,DK02,Wu07}. An interesting issue is to
consider the effect of matter fields on N-P constants. In this
paper, we shall concentrate on the electromagnetic field. Like in
the vacuum case, we require the space-time to be stationary.
Obviously, there is no Bondi energy flux in such a space-time, i.e.
${\dot\sig}^0=0$. In this case we can choose some suitable
coordinates, such that the asymptotic shear $\sigma^0$ is zero.
Similarly, the stationary condition has eliminated the freedom of
the news function.  We also demand the Weyl tensor satisfy the
asymptotically algebraically special condition, which has been
discussed
 above. The main purpose of this paper is to prove the following theorem:
\begin{theorem}
All the N-P constants of an asymptotically flat, stationary,
asymptotically algebraically special electrovacuum space-time are
zero.
\end{theorem}
Note that Kerr-Newman solution satisfies all the conditions in the
theorem. It follows immediately that the all N-P constants in a
Kerr-Newman spacetime must vanish. \\

\noindent {\em Proof of the theorem.}  We choose the standard Bondi-Sachs'
coordinates and construct the standard Bondi null tetrad
\cite{Wu07,Wu06}. With the gauge choice in
\cite{PR86,UN62}, we can write down the N-P coefficients and null
tetrad of the stationary electrovacuum spacetime. Some low order
terms have been calculated and can be found in \cite{PR86}.
Calculation of the N-P constants requires higher order terms in the
expansions. Consider the following N-P equations
\begin{eqnarray}
\de\lambda-\bde\mu=\bar\tau\mu+(\bar\alpha-3\beta)\lambda-\Psi_3+\Phi_{21},\\
\Delta\lambda-\bde\nu=2\alpha\nu+(\bar\gamma-3\gamma-\mu-\bar\mu)\lambda-\Psi_4,
\end{eqnarray}
where  $\Phi_{ij}=8\pi\phi_i\bph_j$ is the Maxwell stress tensor.
The coefficient of  $r^{-2}$  in  equation (2) yields $\Psi_3^0=0$.
Expanding equation (3) up to $O(r^{-3})$, we obtain
$\Psi_4^0=\Psi_4^1=\Psi_4^2=0$.

Now we shall use the Killing equation to reduce other dynamical
freedoms and get a general asymptotic expansion of the stationary
electrovacuum space-time.  Write down the time-like Killing vector
as
\begin{eqnarray}
t^a=Tl^a+n^a+\bA m^a+A\bm^a.\nonumber
\end{eqnarray}
The Killing equations are given by
\begin{eqnarray}
&&-DT+(\gamma+\bga)+\btau A+\tau\bA=0,\label{a2}\\
&&DA+\tau+\brho
A+\sig\bA=0,\label{a3}\\
&&-\De T-(\gamma+\bga)T-\nu A-\bnu\bA=0,\label{a4}\\
&&-\tau T+\bnu+\De A+(\bga-\gamma)A-\de T-\tau T-\mu
A-\blam\bA=0,\label{a5}\\
&&-\sig T+\blam +\de A+(\bal-\be)A=0,\label{a6}\\
&&-\rho T+\mu +\de\bA-(\bal-\be)\bA-\brho T+\bmu +\bde
A-(\al-\bbe)A=0.\label{a7}
\end{eqnarray}
Similarly to the analysis in \cite{Wu07}, assuming the asymptotic
behaviors of $T$ and $A$ as
\begin{eqnarray}
T&=&T^0+\frac{T^1}{r}+\cdots,\nonumber\\
A&=&A^0+\frac{A^1}{r}+\cdots,\label{TA}
\end{eqnarray}
we can solve the Killing equations order by order. The stationary
condition implies ${\dot\sig}^0=0$. It has been found that the
Maxwell field does not change the lowest two powers of $1/r$ in the
Killing equations. So the constant terms in the Killing equations
yield the same result as in the vacuum case, i.e.,  $T^0=\frac12$,
${\dot T}^1=0$, ${\dot A}^1=0$. The coefficients of  $r^{-1}$ in the
Killing equations give rise to $\sig^0=0$, $A^1=0$, ${\dot T}^2=0$,
$\Psi^0_2=\bPsi^0_2$, $T^1=\frac{1}{2}(\Psi^0_2+\bPsi^0_2)$ , ${\dot
A}^2=-\frac{1}{2}\eth\Psi^0_2+\frac{1}{2}\de_0(\Psi^0_2+\bPsi^0_2)$
and
\begin{equation}
{\dot\Psi}^0_2=0.
\end{equation}
From the $r^{-2}$ terms in the N-P equation
\begin{eqnarray}
\delta\nu-\Delta\mu=\gamma\mu-2\nu\beta+\bar\gamma\mu+\mu^2+|\lambda|^2+\Phi_{22}
\,,
\end{eqnarray}\
we find $8\pi|\phi_2^0|^2={\dot\Psi}_2^0=0$. Hence $\phi_2^0=0 $.
From the $r^{-5}$ terms in the N-P equation
\begin{eqnarray}
\delta\rho-\bar\delta\sigma=\tau\rho+(\bar\beta-3\alpha)\sigma+(\rho-\bar\rho)\tau-\Psi_1+\Phi_{01}
\,,
\end{eqnarray}\
we have
\begin{eqnarray}
\frac16(\bbeth\Psi_0^0-40\pi\phi_0^0\bph_1^0)+\frac12\bbeth\Psi_0^0=-\frac13(\bbeth\Psi_0^0-40\pi\phi_0^0\bph_1^0)+\bbeth\Psi_0^0
-16\pi\phi_0^0\bph_1^0
\end{eqnarray}\
which implies
\begin{eqnarray}
\phi_0^0\bph_1^0=0
\end{eqnarray}

This equation will play an important role in our proof, which gives
$\phi_0^0=0$ or $\phi_1^0=0$. Now we discuss the two cases
respectively.

1) $\phi_0^0=0$. Consider the Maxwell equations
\begin{eqnarray}
D\phi_1-\bar\delta\phi_0&=&-2\alpha\phi_0+2\rho\phi_1,\\
D\phi_2-\bar\delta\phi_1&=&-\lambda\phi_0+\rho\phi_2.
\end{eqnarray}
The coefficients of $r^{-4}$ in these equations yield
$\phi_1^1=\phi_2^2=0$.

Consider the other two Maxwell equations
\begin{eqnarray}
\delta\phi_1-\Delta\phi_0&=&(\mu-2\lambda)\phi_0+2\tau\phi_1-\sigma\phi_2,\\
\delta\phi_2-\Delta\phi_1&=&-\nu\phi_0+2\mu\phi_1+(\tau-2\beta)\phi_2.
\end{eqnarray}
The $r^{-2}$ terms in equation (18) and the $r^{-3}$ terms  in
equation (19) yield
\begin{eqnarray}
\dot{\phi_1^0}&=&0\\
\dot{\phi_0^0}&=&\eth\phi_1^0=0\label{case1}
\end{eqnarray}
where ``$\cdot$'' denotes $\ppu$. Combining these two equations, we
have $\phi_1^0=constant$. So from the $r^{-3}$ terms of Eq. (17), we
obtain $\phi_2^1=-\bbeth\phi_1^0=0$.

2)  $\phi_1^0=0$.  Again, from the $r^{-3}$ terms of Eq.(17),  we
have $\phi_2^1=-\bbeth\phi_1^0=0$.

Thus in both cases we have $\phi_2^1=0$. Note that it is
$\Phi_{ij}$, instead of $\phi_i$, that appear in the N-P equations.
The fact that $\phi_i=O(r^{-3})$ (except $\phi_1 \sim O(r^{-2})$  in
case 1)) shows that the presence of the electromagnetic field does
not contribute to $r^{-1}$ and $r^{-2}$ terms.  The electromagnetic
field makes contribution only to order $r^{-3}$ and higher orders in
the expansions. Combining these results, we obtain the reduced N-P
coefficients
\begin{eqnarray}
\rho&=&-\frac{1}{r}+\frac{8\pi\phi^0_0\bar\phi^0_0}{3r^5}+O(r^{-6}),\nonumber\\
\sig&=&-\frac{\Psi^0_0}{2r^4}-\frac{\Psi^1_0}{3r^5}+O(r^{-6}),\nonumber\\
\al&=&\frac{\al^0}{r}-\frac{\bal^0\bPsi^0_0}{6r^4}+\frac{\alpha^08\pi\phi^0_0\bar\phi^0_0
-\bar\alpha^0\bar\Psi^1_0
-24\pi(\phi_1^0\bph_0^1+\phi^1_1\bar\phi^0_0)}{12r^5}+O(r^{-6}),\nonumber\\
\beta&=&-\frac{\bal^0}{r}-\frac{\Psi^0_1}{2r^3}+\frac{\al^0\Psi^0_0
+2\bbeth\Psi^0_0}{6r^4}-\frac{3\Psi^2_1+8\pi\bar\alpha^0\phi^0_0\bar\phi^0_0
-\alpha^0\Psi^1_0}{12r^5}+O(r^{-6})\,,\nonumber\\
\tau&=&-\frac{\Psi^0_1}{2r^3}+\frac{\bbeth\Psi^0_0}{3r^4}+\frac{\bbeth\Psi^1_0-8\pi\eth(\phi^0_0\bar\phi^0_0)
-48\pi(\phi_0^1\bph_1^0+\phi^0_0\bar\phi^1_1)}{8r^5}+O(r^{-6}),\nonumber\\
\lam&=&-\frac{\bPsi^0_0}{12r^4}-\frac{3\bPsi^0_0\Psi^0_2+\bPsi^1_0+48\pi\phi^2_2\bar\phi^0_0}{24r^5}+O(r^{-6}),\nonumber\\
\mu&=&-\frac{1}{2r}-\frac{\Psi^0_2}{r^2}+\frac{\bbeth\Psi^0_1-16\pi\phi_1^0\bph_1^0}{2r^3}-\frac{\bbeth^2\Psi^0_0}{6r^4}
-\frac{6\Psi^3_2+8\pi\phi^0_0\bar\phi^0_0}{24r^5}+O(r^{-6}),\nonumber\\
\gamma&=&-\frac{\Psi^0_2}{2r^2}+\frac{2\bbeth\Psi^0_1-48\pi\phi_1^0\bph_1^0+\al^0\Psi^0_1-\bal^0\bPsi^0_1}{6r^3}\nonumber\\
&&-\frac{1}{24}\left[2\left(\al^0\bbeth\Psi^0_0-\bal^0\eth\bPsi^0_0\right)+3\bbeth^2\Psi^0_0\right]r^{-4}\nonumber\\
&&+\frac1{20}[\al^08\pi(\phi^0_0\bar\phi^1_1+\phi_0^1\bph_1^0)+\al^0\Psi^2_1-\bal^08\pi(\phi_1^0\bph_0^1+\phi_1^1\bar\phi^0_0)
-\bal^0\bar\Psi^2_1\nonumber\\
&&-|\Psi^0_1|^2-4\Psi^3_2-32\pi(\phi_1^0\bph_1^2+\phi_1^1\bar\phi_1^1+\phi_1^2\bph_1^0)]r^{-5}+O(r^{-6}),\nonumber\\
\nu&=&-\frac{1}{12}\left[\bPsi^0_1+2\bbeth^2\Psi^0_1\right]r^{-3}
+\frac{1}{24}\left[\eth\bar\Psi_0^0+\bbeth^3\Psi_0^0\right]r^{-4}\nonumber\\
&&-\frac1{120}[6\Psi_2^1\bar\Psi_1^0-8\Psi_2^0\eth\bar\Psi_0^0+24\pi(\phi_1^0\bar\phi_0^1+\phi_1^1\bar\phi_0^0)+3\bar\Psi_1^2+24\Psi_3^4\nonumber\\
&&+192\pi\phi_2^2\bar\phi_1^1]r^{-5}+O(r^{-6}).\label{connection}
\end{eqnarray}
and the null tetrad
\begin{eqnarray}
l^a&=&\ppr \,,\nonumber\\
n^a&=&\ppu+\left[-\frac{1}{2}-\frac{\Psi^0_2}{r}+\frac{\bbeth\Psi^0_1+\eth\bPsi^0_1+64\pi\phi_1^0\bph_1^0}{6r^2}\right.
-\frac{\bbeth^2\Psi^0_0+\eth^2\bPsi^0_0}{24r^3}\nonumber\\
&&-\frac1{20}\left(3|\Psi_1^0|^2+
\Psi_2^3+\bar\Psi_2^3+16\pi(\phi^0_1\bar\phi^2_1+\phi^1_1\bph^1_1+\phi^2_1\bph^0_1)r^{-4}+O(r^{-5})\right]\ppr\nonumber\\
&&+\left[\frac{1+\zeta\bze}{6\sqrt{2}r^3}\Psi^0_1
-\frac{1+\zeta\bze}{12\sqrt{2}r^4}\bbeth\Psi^0_0+O(r^{-5})\right]\ppz\nonumber\\
&&+\left[\frac{1+\zeta\bze}{6\sqrt{2}r^3}\bPsi^0_1
-\frac{1+\zeta\bze}{12\sqrt{2}r^4}\eth\bPsi^0_0+O(r^{-5})\right]\ppbz\,,\nonumber\\
m^a&=&\left[-\frac{\Psi^0_1}{2r^2}+\frac{\bbeth\Psi^0_0}
{6r^3}-\frac{\Psi_1^2+8\pi(\phi_0^1\bph_1^0+\phi^0_0\bar\phi^1_1)}{12r^4}+O(r^{-5})\right]\ppr\nonumber\\
&&+\left[\frac{1+\zeta\bze}{6\sqrt{2}r^4}\Psi^0_0+O(r^{-5})\right]\ppz
+\left[\frac{1+\zeta\bze}{\sqrt{2}r}+O(r^{-5})\right]\ppbz\,,
\end{eqnarray}
where
$\de_0=\frac{(1+\zeta\bze)}{\sqrt{2}}\frac{\partial}{\partial\bze}$,
$\zeta=e^{i\phi}\cot\frac{\theta}{2}$, $\eth f=(\de_0+2s\bal^0)f$ (
$s$ is the spin-weight of $f$). The differential operators  $\eth$
and $\bbeth$  are defined in \cite{PR86,St90}.

Then the components of the Weyl curvature and the electromagnetic
tensor reduce to
\begin{eqnarray}
&&\Psi_0=\frac{\Psi^0_0}{r^5}+\frac{\Psi^1_0}{r^6}+O(r^{-7}),\nonumber\\
&&\Psi_1=\frac{\Psi^0_1}{r^4}+\frac{\Psi^1_1}{r^5}+\frac{\Psi^2_1}{r^6}+O(r^{-7}),\nonumber\\
&&\Psi_2=\frac{\Psi^0_2}{r^3}+\frac{\Psi^1_2}{r^4}+\frac{\Psi^2_2}{r^5}+\frac{\Psi^3_2}{r^6}+O(r^{-7}),\nonumber\\
&&\Psi_3=\frac{\Psi^2_3}{r^4}+\frac{\Psi^3_3}{r^5}+\frac{\Psi^4_3}{r^6}+O(r^{-7}),\nonumber\\
&&\Psi_4=\frac{\Psi_4^3}{r^4}+\frac{\Psi^4_4}{r^5}+\frac{\Psi^5_4}{r^6}+O(r^{-7}).\nonumber\\
&&\phi_0=\frac{\phi_0^0}{r^3}+\frac{\phi_0^1}{r^4}+\frac{\phi_0^2}{r^5}+O(r^{-6}),\nonumber\\
&&\phi_1=\frac{\phi_1^0}{r^2}+\frac{\phi_1^1}{r^3}+\frac{\phi_1^2}{r^4}+\frac{\phi_1^3}{r^5}+O(r^{-6}),\nonumber\\
&&\phi_2=\frac{\phi_2^2}{r^3}+\frac{\phi_2^3}{r^4}+\frac{\phi_2^4}{r^5}+O(r^{-6}).\label{weyl}
\end{eqnarray}
The Bianchi identity takes the form
\begin{eqnarray}
\bar\delta\Psi_0-D\Psi_1+D\Phi_{01}-\delta\Phi_{00}=4\alpha\Psi_0-
4\rho\Psi_1-2\tau\Phi_{00}+2\rho\Phi_{01}+2\sigma\Phi_{10}\,.
\end{eqnarray}
The coefficient of $r^{-6}$ in equation (25) yields
$\Psi^1_1=-\bbeth\Psi^0_0$.

Similarly, the other components of the  Bianchi identity and the
Maxwell equations lead to
\begin{eqnarray}
&&\phi_1^1=-\bbeth\phi_0^0,\quad \phi_1^2=-\frac12\bbeth\phi_0^1,\quad \phi_1^3=-\frac13\bbeth\phi_0^2-\frac12\bar\Psi_1^0\phi_0^0.\nonumber\\
&&\phi_2^2=\frac12\bbeth^2\phi_0^0,\quad \phi_2^3=\frac16\bbeth^2\phi_0^1,\nonumber\\
&&\phi_2^4=\frac1{12}\bbeth^2\phi_0^2+\frac1{12}\eth\bar\Psi_0^0+\frac12\bar\Psi_1^0\bbeth\phi_0^0\nonumber\\
&&\Psi^1_1=-\bbeth\Psi^0_0,\quad
\Psi^2_1=-\frac{1}{2}\bbeth\Psi^1_0+16\pi(\phi_0^0\bar\phi_1^1+\phi_0^1\bph_1^0)+4\pi\eth(\phi_0^0\bar\phi_0^0),\nonumber\\
&&\Psi^1_2=-\bbeth\Psi^0_1+16\pi\phi_1^0\bph_1^0,\quad \Psi^2_2=\frac{1}{2}\bbeth^2\Psi^0_0,\nonumber\\
&&\Psi^3_2=-\frac{2}{3}|\Psi^0_1|^2-\frac{1}{3}\bbeth\Psi^2_1+\frac{16}9\pi\eth(\phi_1^0\bph_0^1+\phi_1^1\bar\phi_0^0)
-\frac89\pi\bbeth(\phi_0^0\bar\phi_1^1+\phi_0^1\bph_1^0)-\frac{20}{9}\pi\phi_0^0\bar\phi_0^0\nonumber\\
&&\quad\quad\quad+\frac{80}9\pi(\phi_1^0\bph_1^2+\phi_1^1\bph_1^1+\phi_1^2\bph_1^0)+\frac89\pi\ppu(\phi_0^0\bph_0^1+\phi_0^1\bph_0^0),\nonumber\\
&&\Psi^2_3=\frac{1}{2}\bbeth^2\Psi^0_1,\quad \Psi^3_3=-\frac{1}{2}\bPsi^0_1\Psi^0_2-\frac{1}{6}\bbeth^3\Psi^0_0,\nonumber\\
&&\Psi^4_3=-\frac14\bbeth\Psi_2^3+\frac18\Psi_2^0\eth\Psi_0^0+\frac12\bPsi_1^0\bbeth\Psi_1^0+\frac1{12}k\eth(\phi_2^2\bph_0^0)\nonumber\\
&&\quad\quad\quad-\frac43\pi\bbeth(\phi_1^0\bph_1^2+\phi_1^1\bph_1^1\phi_1^2\bph_1^0)+4\pi(\phi_2^2\bph_1^1+\phi_2^3\bph_1^0)
+4\pi(\phi_1^0\bph_0^1+\phi_1^1\bph_0^0)\nonumber\\
&&\quad\quad\quad+4\pi\bar\Psi_1^0\phi_1^0\bph_1^0+\frac43\pi\ppu(\phi_1^1\bph_0^1+\phi_1^2\bph_0^0),\nonumber\\
&&\Psi^3_4=-\frac16\bbeth^3\Psi_1^0,\quad \Psi^4_4=-\frac1{24}\bbeth^4\Psi_0^0,\nonumber\\
&&\Psi^5_4=-\frac15\bbeth\Psi_3^4-\frac85\pi\bbeth(\phi_2^2\bph_1^1+\phi_2^3\bph_1^0)-\frac15\bPsi_1^0\bbeth^2\Psi_1^0
-\frac1{20}\Psi_2^0\bar\Psi_0^0\nonumber\\
&&\quad\quad\quad+4\pi(\phi_2^2\bph_0^0)+\frac85\pi\ppu(\phi_2^2\bph_0^1+\phi_2^3\bph_0^0).\label{weyl2}
\end{eqnarray}

Similarly to the treatment in \cite {Wu07},  the $r^{-3}$ terms in
the  Killing equations lead to $\eth\Psi_1^0=0$. Thus we have
\begin{eqnarray}
\Psi^0_1&=&\sum^1_{m=-1}B_m{\ }_1Y_{1,m},\nonumber\\
\Psi^0_2&=&C.\label{aaa}
\end{eqnarray}

The coefficient of $r^{-3}$ in Eq. (2) gives
$\Psi_3^1=\bde\Psi_2^0=0$.

In order to find more restrictions on $\Psi_0$, we need to compute
higher order terms of the Killing equations. The terms of order
$r^{-4}$ of the Killing equations yield
\begin{eqnarray}
&&3T^3+(\gamma^4+\bga^4)=0,\label{24}\\
&&4A^3=\frac{1}{3}\bbeth\Psi^0_0,\label{34}\\
&&{\dot T}^4+\frac83\pi\phi_1^0\bph_1^0=0,\label{44}\\
&&\frac{1}{2}\Psi^0_1T^1-\tau^4+\bnu^4+{\dot
A}^4+(\Psi^0_2+\bPsi^0_2)A^2+2A^3-\de_0 T^3+\Psi^0_2A^2=0,\label{54}\\
&&\frac{1}{6}\Psi^0_0+\eth A^3=0,\label{64}\\
&&2T^3+\mu^4+\bmu^4+\eth\bA^3+\bbeth A^3=0 \,.\label{74}
\end{eqnarray}
Eq.(\ref{34}) and (\ref{64}) imply
\bee
\Psi^0_0=\sum^2_{m=-2}A_m(u){\ }_2Y_{2,m},\label{Psi00} \ede Eq.(27)
and ${\dot T}^3=0$ ( which comes from the $r^{-3}$ terms in the
Killing equations) imply that $\Psi^0_0$ is independent of $u$.

Combining  Eqs.(\ref{weyl}),(\ref{weyl2}),(\ref{aaa}) and
(\ref{Psi00}), one finds
\begin{eqnarray}
I^3-27J^2\sim O(r^{-21}).
\end{eqnarray}
This result holds for a general asymptotically flat stationary
spacetime. As mentioned in the introduction, the AASC requires
\begin{eqnarray}
I^3-27J^2\sim O(r^{-22}),
\end{eqnarray}
which is just one order faster than the falloff rate of a general
asymptotically flat spacetime. This means that the AASC is a  weak
requirement and as demonstrated at the end of this section, there
exist many asymptotic flat space-times which satisfy this condition.

Our purpose is to calculate the Newman-Penrose constants, which are
contained in the coefficients of $\Psi^1_0$.  From the $r^{-5}$
terms in the Killing equations, we have
\begin{eqnarray}
&&4T^4+(\gamma^5+\bga^5)-\frac{1}{2}\bPsi^0_1A^2-\frac{1}{2}\Psi^0_1\bA^2=0,\label{25}\\
&&A^4=\frac{1}{5}\tau^5=\frac1{40}\left[\bbeth\Psi_0^1-48\pi(\phi_0^0\bph_1^1+\phi_0^1\bph_1^0)-8\pi\eth(\phi_0^0\bph_0^0)\right],\label{35}\\
&&\frac18\Psi_0^1+\frac38\Psi_0^0\Psi_2^0-2\pi\phi_0^0\bph_2^2-\frac14(\Psi_1^0)^2+\eth A^4=0,\label{65}\\
&&-\rho^5+2T^4+(\mu^5+\bmu^5)+\frac{3}{2}\Psi^0_1\bA^2+\frac{3}{2}\bPsi^0_1A^2
+\eth\bA^4+\bbeth A^4=0 \,.\label{75}
\end{eqnarray}
Eqs. (\ref{35}) and (\ref{65}) yield:
\begin{eqnarray}
\eth\bbeth\Psi^1_0+5\Psi^1_0=10(\Psi^0_1)^2-15\Psi^0_0\Psi^0_2+80\pi\phi_0^0\bph_2^2+48\pi\eth(\phi_0^0\bph_1^1+\phi_0^1\bph_1^0)
+8\pi\eth^2(\phi_0^0\bph_0^0).\label{Psi01}
\end{eqnarray}
The terms of $\phi^i_j$ on the right-hand side of \eq{Psi01} are the
contribution from the Maxwell field
\cite{Wu07}. To simplify this equation, we need to
investigate the electromagnetic field in more detail.

Since the electromagnetic field is stationary, we have
$\pounds_{t}F_{ab}=0$, where $t^c$ is the Killing vector. Noting
that $\phi_0=F_{lm}$ and using the expansion of $t^c$, we have
\begin{eqnarray}
\pounds_{t}\phi_0&=&\pounds_{t}F_{ab}l^am^b\nonumber\\
&=&(Tl^c+n^c+\bA m^c+A\bm^c)\phi_0\nonumber\\
&=&F_{ab}l^a[t,\quad m]^b+F_{ab}m^b[t,\quad l]^a\nonumber\\
&=&(\gamma+\bar\gamma+\bar A\bar\tau+A\tau)\phi_0-(\tau+\bar A\sigma+A\rho)(\phi_1-\bph_1)\nonumber\\
&&
+\left[T\bar\varrho-\mu+\gamma+\bar\gamma-A(\bar\beta-\alpha)\right]\phi_0+\left[T\sigma-\bar\lambda-A(\bar\alpha-\beta)\right]\bph_0
\end{eqnarray}
where $[t,\  m]^b$ denotes the commutator of $t^c$ and $m^b$. So we
obtain
\begin{eqnarray}
&&(Tl^c+n^c+\bA m^c+A\bm^c)\phi_0 \nonumber \\
&=&(\gamma+\bar\gamma+\bar A\bar\tau+A\tau)\phi_0 -(\tau+\bar
A\sigma+A\rho)(\phi_1-\bph_1)
+\left[T\bar\varrho-\mu+\gamma+\bar\gamma-A(\bar\beta-\alpha)\right]\phi_0\nonumber\\
&&+\left[T\sigma-\bar\lambda-A(\bar\alpha-\beta)\right]\bph_0
\label{killing}
\end{eqnarray}
Substituting (23) into (\ref{killing}) yields:
\begin{eqnarray}
&&\ppu\phi_0+\left[-\frac{1}{2}-\frac{\Psi^0_2}{r}+O(r^{-2})\right]\ppr\phi_0
+\left[\frac{1+\zeta\bze}{6\sqrt{2}r^3}\Psi^0_1+O(r^{-4})\right]\ppz\phi_0\nonumber\\
&&+\left[\frac{1+\zeta\bze}{6\sqrt{2}r^3}\bPsi^0_1+O(r^{-4})\right]\ppbz\phi_0
+T\ppr\phi_0-\bar A\left[\frac{\Psi^0_1}{2r^2}+O(r^{-3})\right]\ppr\phi_0\nonumber\\
&&+\bar
A\left[\frac{1+\zeta\bze}{6\sqrt{2}r^4}\Psi^0_0+O(r^{-5})\right]\ppz\phi_0
+\bar
A\left[\frac{1+\zeta\bze}{\sqrt{2}r}+O(r^{-5})\right]\ppbz\phi_0\nonumber\\
&&-A\left[\frac{\bPsi^0_1}{2r^2}+O(r^{-3})\right]\ppr\phi_0\nonumber\\
&&+A\left[\frac{1+\zeta\bze}{6\sqrt{2}r^4}\bPsi^0_0+O(r^{-5})\right]\ppbz\phi_0
+ A\left[\frac{1+\zeta\bze}{\sqrt{2}r}+O(r^{-5})\right]\ppz\phi_0\nonumber\\
&=&(\gamma+\bar\gamma+\bar A\bar\tau+A\tau)\phi_0-(\tau+\bar A\sigma+A\rho)(\phi_1-\bph_1)\nonumber\\
&&+\left[T\bar\varrho-\mu+\gamma+\bar\gamma-A(\bar\beta-\alpha)\right]\phi_0
+\left[T\sigma-\bar\lambda-A(\bar\alpha-\beta)\right]\bph_0
\label{killing2}
\end{eqnarray}

Again,  we compute the $\phi^i_j$ terms in Eq.(\ref{Psi01}) in the
two cases.

For case 1) $\phi_0^0=0$, computing the coefficient of $r^{-5}$ in
Eq. (\ref{killing2}), we obtain
\begin{eqnarray}
\dot{\phi_0^2}=-3\Psi_2^0\phi_0^0=0
\end{eqnarray}
The coefficient of $r^{-5}$ of  equation (18) gives
\begin{eqnarray}
\eth\phi_1^1-\dot{\phi_0^2}-2\phi_0^1-3\Psi_2^0\phi_0^0=-\frac12\phi_0^1-\Psi_1^0\phi_1^0
\,.
\end{eqnarray}
Using $\phi_0^0=0$ and $\dot{\phi_0^2}=0$, we get
\begin{eqnarray}
\phi_0^1=\frac23\phi_1^0\Psi_1^0 \,.
\end{eqnarray}
By taking  $\eth$ on both sides and using $\eth\Psi_1^0=0$, we have
immediately
\begin{eqnarray}
\eth\phi_0^1=\frac23\phi_1^0\eth\Psi_1^0=0\label{case12}.
\end{eqnarray}

Then the $\phi^i_j$ terms in Eq.(\ref{Psi01}) become
\begin{eqnarray}
&&80\pi\phi_0^0\bph_2^2+48\pi\eth(\phi_0^0\bph_1^1+\phi_0^1\bph_1^0)
+8\pi\eth^2(\phi_0^0\bph_0^0)\nonumber\\
&=&48\pi\eth(\phi_0^1\bph_1^0)\nonumber\\
&=&(48\pi\eth\phi_0^1)\bph_1^0+\phi_0^1(48\pi\eth\bph_1^0)\nonumber\\
&=&0 \,,
\end{eqnarray}
where \eqs{case1} and \meq{case12} have been used in the last step.

For case 2) $\phi_1^0=0$,  the coefficient of $r^{-4}$ in Eq.
(\ref{killing2}) leads to
\begin{eqnarray}
0=\dot{\phi_0^1}-3T^0\phi_0^0+\frac32\phi_0^0=\dot{\phi_0^1} \,.
\label{phi01}
\end{eqnarray}

Because the spinweight of $\phi_0$ is 1, we can expand $\phi_0^0$ as
$\phi_0^0=\sum_{l=1}^{\infty}\sum_{m=-l}^l d_{l,m}{\ }_1Y_{l,m}$,
where $d_{l,m}$ are some constants. The $r^{-4}$ terms in (18) yield
\begin{eqnarray}
\dot{\phi_0^1}=-\bbeth\eth\phi_0^0=\frac12\sum_{l=1}^{\infty}(l+2)(l-1)\sum_{m=-l}^{l}d_{l,m}
\ {}_1Y_{l,m} \,. \label{dphi}
\end{eqnarray}
Combining (\ref{phi01}) and (\ref{dphi}) and using the fact that
spin-weight harmonic function components are linearly independent,
we obtain $l=1$. Consequently,
\begin{eqnarray}
\phi_0^0=\sum_{m=-1}^1 d_m{\ }_1Y_{1,m} \,,
\end{eqnarray}
where $d_m$ are constants. By expanding  $\phi_0^0$, we find
$\eth\phi_0^0=0$. The contribution from the Maxwell field in Eq.
(\ref{Psi01}) then leads to:
\begin{eqnarray}
&&80\pi\phi_0^0\bph_2^2+48\pi\eth(\phi_0^0\bph_1^1)+8\pi\eth^2(\phi_0^0\bph_0^0)\nonumber\\
&=&40\pi\phi_0^0\eth^2\bph_0^0-48\pi\eth(\phi_0^0\eth\bph_0^0)+8\pi\eth(\phi_0^0\eth\bph_0^0+\bph_0^0\eth\phi_0^0)\nonumber\\
&=&40\pi\phi_0^0\eth^2\bph_0^0-48\pi\eth\phi_0^0\eth\bph_0^0-48\pi\phi_0^0\eth^2\bph_0^0
+8\pi\eth\phi_0^0\eth\bph_0^0\nonumber\\
&&+8\pi\eth\bph_0^0\eth\phi_0^0+8\pi\phi_0^0\eth^2\bph_0^0+8\pi\bph_0^0\eth^2\phi_0^0\nonumber\\
&=&-32\pi\eth\phi_0^0\eth\bph_0^0+8\pi\bph_0^0\eth^2\phi_0^0\nonumber\\
&=&0
\end{eqnarray}
where we have used $\phi_1^1=-\bbeth\phi_0^0$ and
$\phi_2^2=\frac12\bbeth^2\phi_0^0$. Therefore, the electromagnetic
field makes no contribution to the equation of $\Psi_0^1$. So in
either case, the equation of $\Psi_0^1$ reduces to
\begin{eqnarray}
\eth\bbeth\Psi^1_0+5\Psi^1_0=10(\Psi^0_1)^2-15\Psi^0_0\Psi^0_2 \,,
\label{fieq}
\end{eqnarray}
which is exactly the same equation as that in the vacuum case. Then
by imposing the AASC, it is shown  in \cite{Wu07} that \eq{fieq}
implies that all the Newman-Penrose constants must be zero. This
completes the proof of our theorem.

Remark : The asymptotically algebraically special condition has
played an important role in the proof of this paper and in
\cite{Wu07}.
Obviously, this condition is satisfied by the Kerr-Newman solution.
The following arguments show that the AASC is a rather weak
condition imposed on a general asymptotically flat spacetime. Note
that the Kerr-Newman spacetime is axisymmetric. Such symmetry is not
required in our theorem. From Eq.(\ref{aaa}), we can see that
$\Psi^0_1$ contains ${}_1Y_{1,1}$ and ${}_1Y_{1,-1}$ components that
do not appear in the Kerr-Newman solution. Simple calculation shows
that $span\{{}_1Y_{1,1},{}_1Y_{1,0},{}_1Y_{1,-1}\}$ is not a
representative space of $SO(3)$. Thus  we cannot cancel such
components by a rotation. Based on the characteristic initial value
method\cite{Fr81}, it is not difficult to construct exact solutions
with non-zero $B_1$ and $B_{-1}$. Furthermore, the spin-weight
components of $\Psi^k_0$ are just the Janis-Newman multi-poles of
gravitational field\cite{JN65}. The AASC only gives a restriction
between Janis-Newman's dipoles and quadrupoles\cite{Wu07}. Since
there is no restriction on higher order multi-poles, it is easy to
see that there are many solutions which satisfy the conditions of
our theorem and are not equivalent to the Kerr-Newman solution.

\section{Concluding remarks}
We have proven that all the N-P constants of an asymptotic flat,
stationary, asymptotically algebraically special electrovacuum
space-time are zero. The Kerr-Newman solution  manifestly satisfies
all the conditions. So our theorem implies that  all the N-P
constants of the Kerr-Newman solution are zero. This result has been
obtained resently\cite{Gong} by other authors. In the proof of the
theorem, we have assumed that the Maxwell field is stationary. If
this condition is not imposed, ${\dot\phi}^1_0$ will not be zero.
Then Eq.(\ref{dphi}) tells us $\phi^0_0$ will contain other
components of the spin-weight spherical functions. These terms
correspond to the Janis-Newman multi-pole of Maxwell
field\cite{JN65}. In the presence of these terms, the N-P constants
may not vanish. Last but not least, an interesting issue is to
single out the Kerr-Newman solution from solutions which satisfy the
conditions of our theorem. From the discussion of the last section,
we find that the AASC is not enough to uniquely determine the
Kerr-Newman solution. It seems that more restrictions on the Maxwell
field are needed. This will be discussed in our future work.

\section*{Acknowledgement}
This work is supported by the Natural Science Foundation of China
(NSFC) under Grant Nos. 10705048, 10605006, 10731080. Authors would
like to thank the referees for helpful comments on the
asymptotically algebraically special condition.

\end{document}